\documentclass[10pt,twocolumn,letterpaper]{article}

\usepackage[pagenumbers]{cvpr} 

\usepackage{graphicx}
\usepackage{amsmath}
\usepackage{amssymb}
\usepackage{booktabs}
\usepackage{fancyhdr}

\usepackage[pagebackref,breaklinks,colorlinks]{hyperref}

\usepackage{tabularx}
\usepackage{booktabs}
\usepackage{multirow}
\usepackage{algpseudocode}
\usepackage{algorithm}

\usepackage{amssymb}
\usepackage{caption}
\usepackage{subcaption}

\usepackage[accsupp]{axessibility}  

\usepackage{subcaption}
\usepackage[capitalize]{cleveref}
\crefname{section}{Sec.}{Secs.}
\Crefname{section}{Section}{Sections}
\Crefname{table}{Table}{Tables}
\crefname{table}{Tab.}{Tabs.}

\def\cvprPaperID{11527} 
\def\confName{CVPR}
\def\confYear{2023}

\begin{document}

\renewcommand{\headrulewidth}{0pt}
\addtolength{\topmargin}{-12.0pt}
\setlength{\headheight}{12pt}
\fancyhf{}
\fancyhead[C]{To appear in CVPR 2023}

\title{Jedi: Entropy-based Localization and Removal of Adversarial Patches}

\author{Bilel Tarchoun \textsuperscript{$\ast$},
Anouar Ben Khalifa \textsuperscript{$\ast,\perp$},
Mohamed Ali Mahjoub  \textsuperscript{$\ast$},
Nael Abu-Ghazaleh  \textsuperscript{$\dagger$}
\and 
Ihsen Alouani \textsuperscript{$\ddagger, \Upsilon$}\\
\textsuperscript{$\ddagger$} CSIT, Queen's University Belfast, UK \\
\textsuperscript{$\ast$} Université de Sousse, Ecole Nationale d’Ingénieurs de Sousse, LATIS, Sousse, Tunisia; \\
\textsuperscript{$\Upsilon$}IEMN CNRS 8520, Université Polytechnique Hauts-de-France\\
\textsuperscript{$\perp$} Université de Jendouba, Institut National des Technologies  et des Sciences du Kef, Tunisia; \\
\textsuperscript{$\dagger$} University of California Riverside, CA, USA \\
{\tt\small bilel.tarchoun@eniso.u-sousse.tn,i.alouani@qub.ac.uk}
}
\maketitle
\thispagestyle{fancy}
\begin{abstract}
Real-world adversarial physical patches were shown to be successful in compromising state-of-the-art models in a variety of computer vision applications.
Existing defenses that are based on either input gradient or features analysis have been compromised by recent GAN-based attacks that generate naturalistic patches. 
In this paper, we propose {\em Jedi}, a new defense against adversarial patches that is resilient to realistic patch attacks.  Jedi tackles the patch localization problem from an information theory perspective; leverages two new ideas: (1) it improves the identification of potential patch regions using \textbf{entropy analysis}: we show that the entropy of adversarial patches is high, even in naturalistic patches; and (2) it improves the localization of adversarial patches, using an autoencoder that is able to complete patch regions from high entropy kernels.  
Jedi achieves high-precision adversarial patch localization, which we show is critical to successfully repair the images. Since Jedi relies on an input entropy analysis, it is \emph{model-agnostic}, and can be applied on pre-trained off-the-shelf models without changes to the training or inference of the protected models. Jedi detects on average $90 \%$ of adversarial patches across different benchmarks and recovers up to $94 \%$ of successful patch attacks (Compared to 75\% and 65\% for LGS and Jujutsu, respectively). 

\end{abstract}

\section{Introduction}
\label{sec:intro}
Deep neural networks (DNNs) are vulnerable to adversarial attacks \cite{szegedy_noise} where an adversary adds carefully crafted imperceptible perturbations to an input (e.g., by $l_p$-norm bounded noise magnitude), forcing the models to misclassify. Several adversarial noise generation methods have been proposed ~\cite{FGSM,cw}, often as part of a cat and mouse game where new defenses emerge\cite{noise_def1,noise_def2} only to be shown vulnerable to new adaptive attacks. 
\begin{figure*}[tp]
  \centering
  \includegraphics[width=1.9\columnwidth]{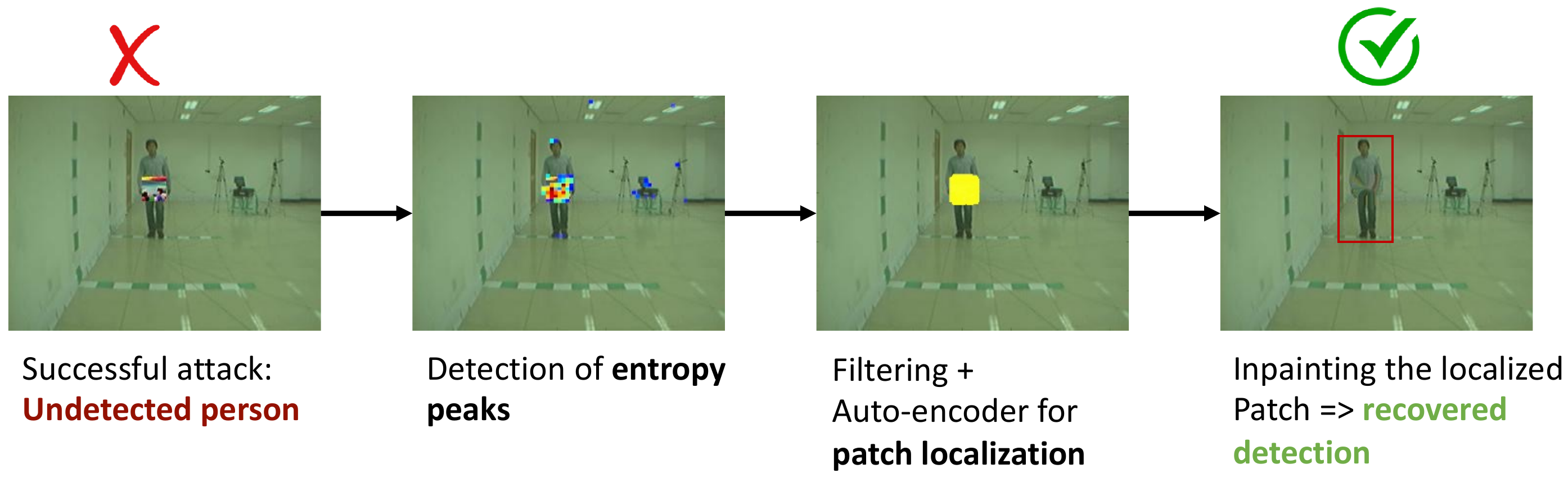}
  \caption{Illustration on a patch attack against Yolo. Jedi surgically localizes the patch via entropy analysis and recovers the image.}
  \label{fig_3_1}
\end{figure*}
Under real-world conditions, an attacker creates a physical patch (thus, spatially constrained) that contains an adversarial pattern.  Such a patch can be placed as a sticker on traffic signs  \cite{evt_stop}, worn as part of an item of clothing \cite{yolopatch,naturalistic}, or introduced using a display monitor \cite{tvpatch}, providing a practical approach for attackers to carry out adversarial attacks.   
 First proposed by Brown et al. \cite{advpatch}, these patches are different from traditional adversarial attacks in two primary ways: (1) they occupy a constrained space within an image; and (2) they may not be noise budget-constrained within the patch. Several adversarial patch generation methods have been demonstrated \cite{yolopatch,lavan,dpatch,naturalistic}, many of which showcasing real-life implementation, making them an ongoing threat to visual ML systems.

Several approaches aiming to detect adversarial patches and defuse their impact have been proposed~\cite{lgs,dw,lance,jujutsu,derandomized,patchguard,patchguardpp}. 
One category of defenses attempts to locate patches by detecting anomalies caused by the presence of the patch. These anomalies can be identified in the input pixel data such as in the case of Localized Gradient Smoothing \cite{lgs} where the patch is located using high pixel gradient values.  Alternatively, they can be identified in the feature space where the adversarial patch can create irregular saliency maps with regards to its targeted class that can be exploited by defenses such as in Digital Watermarking \cite{dw} and Jujutsu \cite{jujutsu}.  
These defenses have two primary limitations: (1) They are only moderately successful against baseline attacks enabling recovery from many attacks (e.g., $75\%$ for LGS and $65\%$ for Jujutsu); and (2) they are vulnerable to adaptive attacks that generate {\em naturalistic} adversarial patches that are meant to use patterns similar to natural images. Hu et al.~\cite{naturalistic} train a GAN to generate naturalistic patches, matching the visual properties of normal images, and show that they are able to bypass defenses that are based on either input or features analysis. 

In this paper, we propose {\em Jedi}\footnote{We use the name Jedi, because the system recognizes chaos (high entropy patches) and restores peace by removing them, like the Jedis in the Star Wars movies.}, a new defense approach that surgically localizes adversarial patches to finally reconstruct the initial image.   Compared to state of the art, {\em Jedi} enables both more accurate detection and recovery from patches, as well as resilience to adaptive attacks leveraging two primary ideas: (1) Using \textbf{entropy} to improve the identification of suspicious regions of the image: we show that entropy can serve as an excellent discriminator of likely patch regions.  Importantly, we also show that differential entropy analysis detects effectively even when naturalistic adaptive attacks are applied; (2) More accurate localization of patches using a trained auto-encoder: we show that a primary reasons behind the limited effectiveness of prior solutions is their inaccurate localization of patches. We substantially improve patch localization, raising the rate of accurately located patches (IoU $> 0.5$) to twice compared to other related approaches. Moreover, Jedi improves the recovery rate while reducing by half the lost detection rate compared to the other evaluated adversarial patch defenses. 

We conduct comprehensive experiments under different datasets and for different scenarios. Our results show that Jedi localizes adversarial patches with high precision, which consequently leads to an effective patch mitigation process, restoring up to $94 \%$ of incorrect results caused by adversarial patch attacks.  More importantly,  Jedi remains efficient with up to $76\%$ recovery rate against GAN-based Naturalistic Patch \cite{naturalistic}, which almost completely bypasses other defenses. Besides, we propose a new adaptive attack that comprehensively limit entropy of the generated patch in Section \ref{sec:adapt}. We find that it is difficult to limit adversarial noise entropy without losing the patch efficiency. As a result, we believe that entropy is a strong feature to discriminate between adversarial patches and benign images. 

Our contributions can be summarized as follows:




\noindent \textbf{i)} We propose Jedi, an entropy-based approach to defend against adversarial patches. Jedi leverage an entropy analysis based approach for a precise patch localization, which results in a high recovery success using inpainting. To our knowledge, we are first to use entropy for adversarial patch localization.

\noindent \textbf{ii)} We evaluate our defense in a variety of settings for both classification and detection tasks and find that Jedi recovers up to $94 \%$ of successful adversarial patch attacks
    
\noindent \textbf{iii)} We propose an entropy-aware adaptive attack that considers entropy budget in the patch generation process. We show that limiting patch entropy escapes detection, but reduces considerably the patches' adversarial impact. 

\noindent \textbf{iv)} For reproducible research, we open source our code (provided in supplementary material with a demo video).



%
\section{Preliminaries: Entropy Analysis}\label{sec:observ}

We provide a preliminary analysis from an information theory perspective to investigate entropy's discriminatory potential between natural images and adversarial patches. 

\noindent\textbf{Predictability in natural images.} 
One measure of an image's \textit{non-randomness} is the level of predictability: if one has access to parts of a given image, what is the capacity to guess the missing parts. If it is composed of totally random pixels, there is no predictability. However, natural images have semantic long-range correlations, and hence a high predictability is expected.

\noindent\textbf{High entropy in adversarial patches.} 
Adversarial patches are, by definition, not natural; they are the result of solving a constrained optimization problem. This adversarial perturbation is designed to be universal, which means that one specific noise is designed to fool a model for a variety of inputs. Moreover, the real-life settings of the threat model considered in adversarial patches represent additional spatial constraints on the adversarial noise generation. In fact, to build a plausible real-life attack, the noise has to be limited by a specific location and cannot be distributed all over the input. Therefore, while natural images have a form of semantic continuity, adversarial patches concentrate a high amount of information within a restricted area, resulting in a highly \textit{unpredictable} image.


Our intuition based on this observation is that adversarial patches should contain a statistically higher amount of information, from an information theory perspective, compared to any random neighborhood from a natural image distribution. Therefore, we use Shannon's entropy as an indicator of adversarial patch candidates. To verify this intuition, we propose a preliminary study to investigate the statistical components of different adversarial patches generated for several models. We also observe the patch entropy behavior in the adversarial noise generation process, comparatively with the mean entropy of the corresponding clean data distribution.

First, we propose to compare the statistical distribution of localized entropy levels in natural images and adversarial patches. The natural images used for this study come from cropping random $50 \times 50$ pixel areas from sample images in the datasets used for our experiments in Section \ref{sec:res}. As for the adversarial patches, we create a collection of adversarial perturbations, as well as patches featured in other adversarial attack papers. Similarly to natural images, we use a sliding window to create $50 \times 50$ pixel sub-images. 

\begin{figure}
    \centering
    \begin{subfigure}[t]{0.495\columnwidth}
        \centering
        \includegraphics[width=\textwidth]{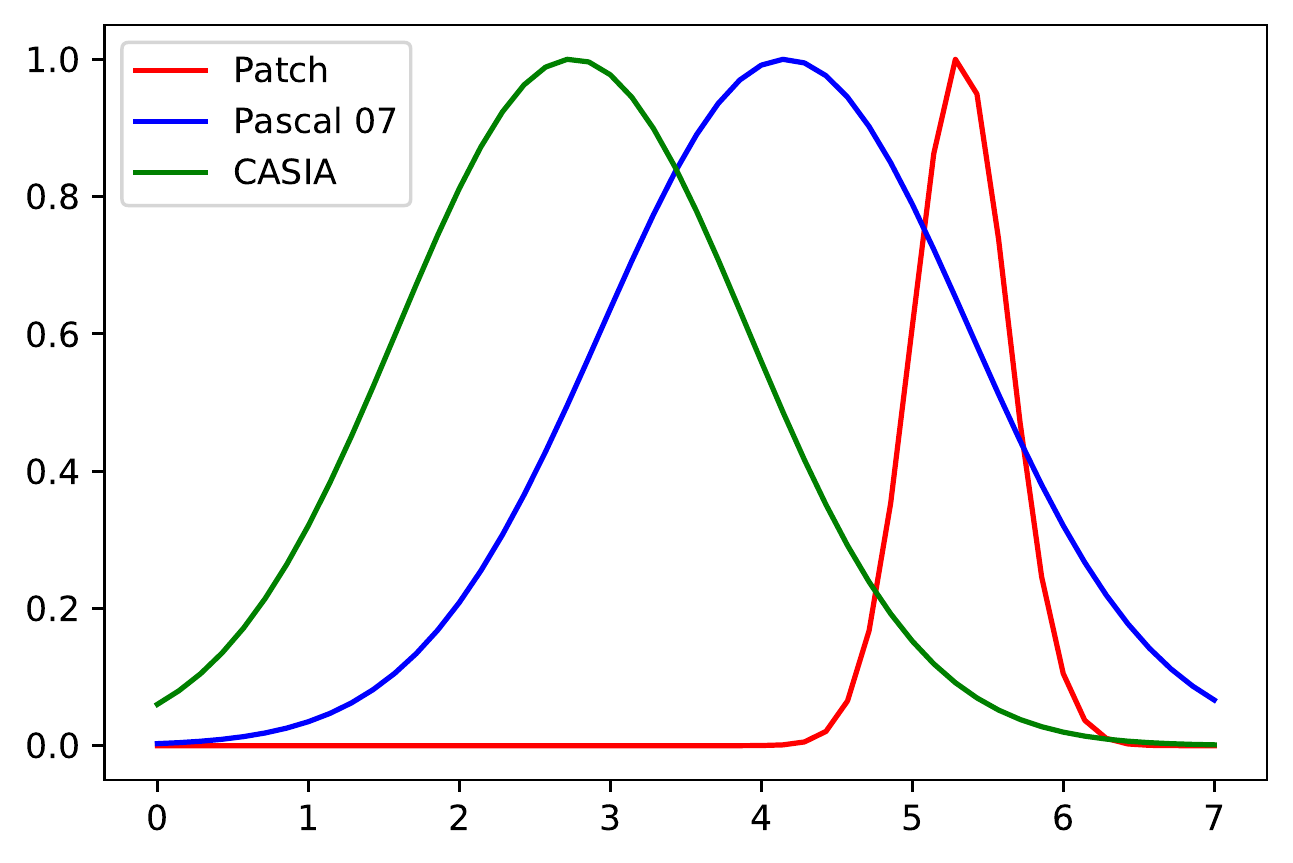}
        \caption{}
        \label{fig_2_1}
    \end{subfigure}
    \begin{subfigure}[t]{0.495\columnwidth}
        \centering
        \includegraphics[width=\textwidth]{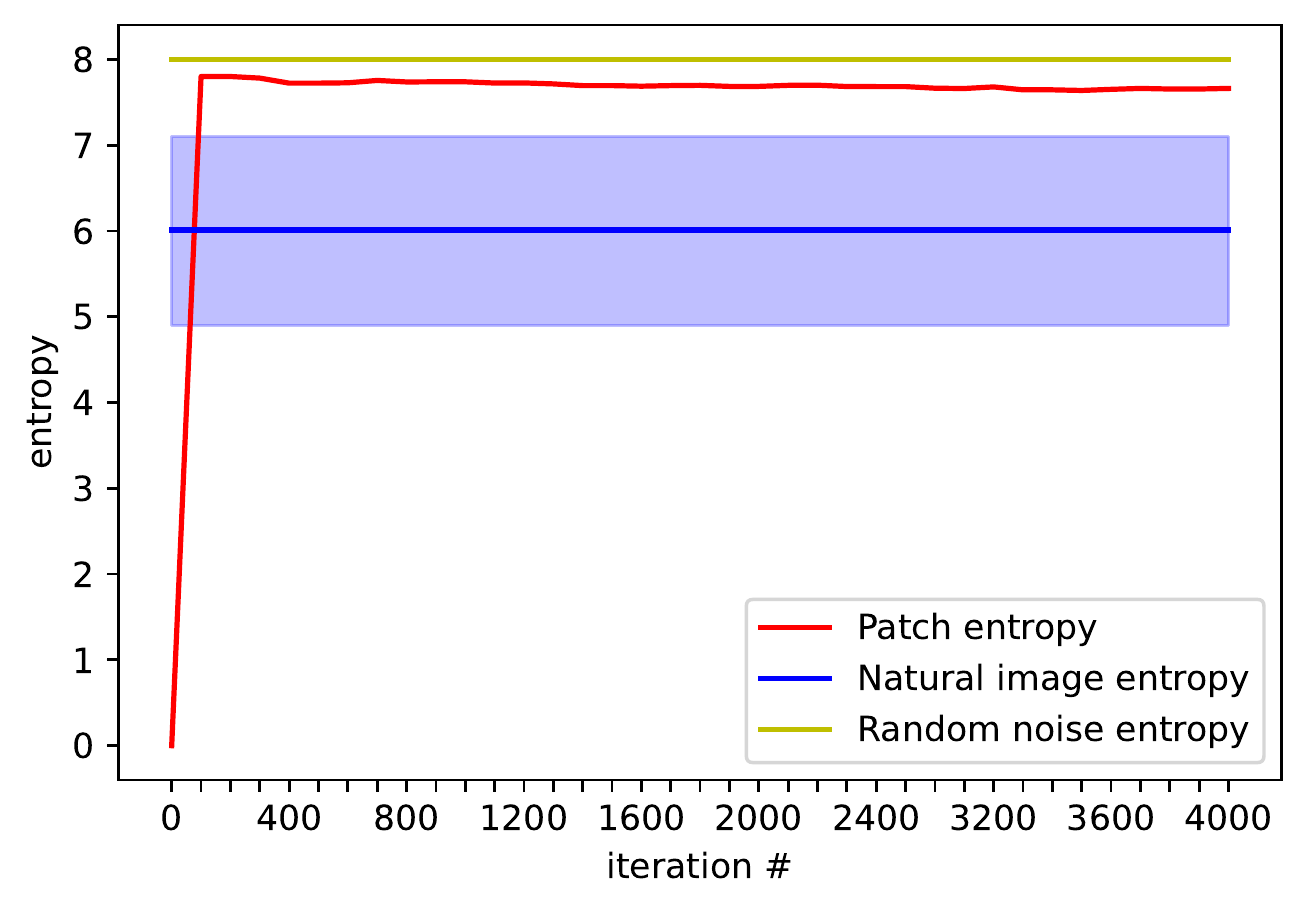}
        \caption{}
        \label{fig_2_2}
    \end{subfigure}
    \caption{a) Comparison of entropy distributions of natural images from Pascal VOC 07 and CASIA datasets and adversarial patches. b)Evolution of patch entropy during the generation process}
    \label{fig_entr_analysis}
\end{figure}

Figure \ref{fig_2_1} shows a considerable entropy distribution shift with adversarial patches having approximately $30\%$ higher mean entropy than natural images, at least. This shift is significant enough to be exploited as a metric to distinguish adversarial patches from natural images. However, the diverse environments and sparsity of natural images result in an entropy distribution with a large standard deviation, which in turn results in a slight overlap between both distributions. Therefore, a comprehensive entropy-based discrimination is required to avoid false positives.

To further explore the noise behavior at design time, we study the evolution of entropy levels in the adversarial patch during the patch generation process. We run the patch generation proposed in \cite{advpatch} while monitoring its entropy. The results are depicted in Figure \ref{fig_2_2}, along with: (1) the mean and standard deviation of the entropy distribution of natural images of the same size cropped from the Pascal VOC 07 dataset, and (2) the entropy of a totally random image representing the theoretical maximum entropy level for this signal window.
The starting point for the patch generation process is a uniform image. With the noise exploration progress, the entropy quickly rises to exceed the mean entropy of similar size natural images, and is close to the theoretical maximum entropy. This suggests that effective patches trend towards higher entropy as their effectiveness rises, with entropy values being comparable to random noise.



\section{Proposed Approach}
\label{sec:prop}
The differential entropy analysis in Section \ref{sec:observ} represents the basis of Jedi to locally discriminate adversarial patches from their surrounding natural image data. Our framework, as illustrated in Figure \ref{fig_3_1}, is based on 3 main steps: 

\subsection{Locating high entropy kernels}
Adversarial patches are likely located in the high entropy clusters. The first step of our approach is to identify these clusters by building a heat map of \emph{local entropy}. A sliding window is applied through the image, where the local Shannon's entropy of the window is calculated. The resulting local entropy values form an entropy heat map. To keep only high entropy kernels in the heat map, a thresholding operation is applied. 

\begin{figure}[tp]
  \centering
  \includegraphics[scale = 0.6]{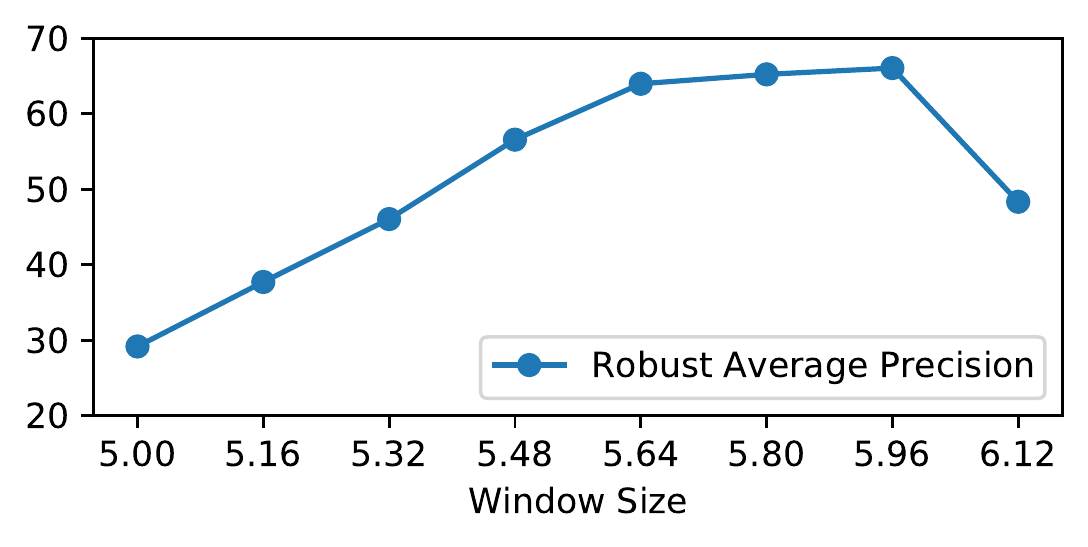}
  \caption{Robust Accuracy for different static entropy thresholds }
  \label{fig_prop_static}
\end{figure}

\noindent\textbf{Threshold:} An appropriate threshold is key to properly isolating the high entropy kernels that constitute potential patch candidates. A too high threshold leads to the inability to detect adversarial patches, while a too low threshold results in a high false positive rate. As Figure \ref{fig_prop_static} shows, the evolution of the robust accuracy as the entropy threshold lowers suggests that using a static threshold is not an appropriate choice. In fact, uncontrolled settings such as outdoor environments have significant variation, which makes a threshold for one image not necessarily fit for another image.
 \begin{figure}[!tp]
  \centering
  \includegraphics[scale = 0.5]{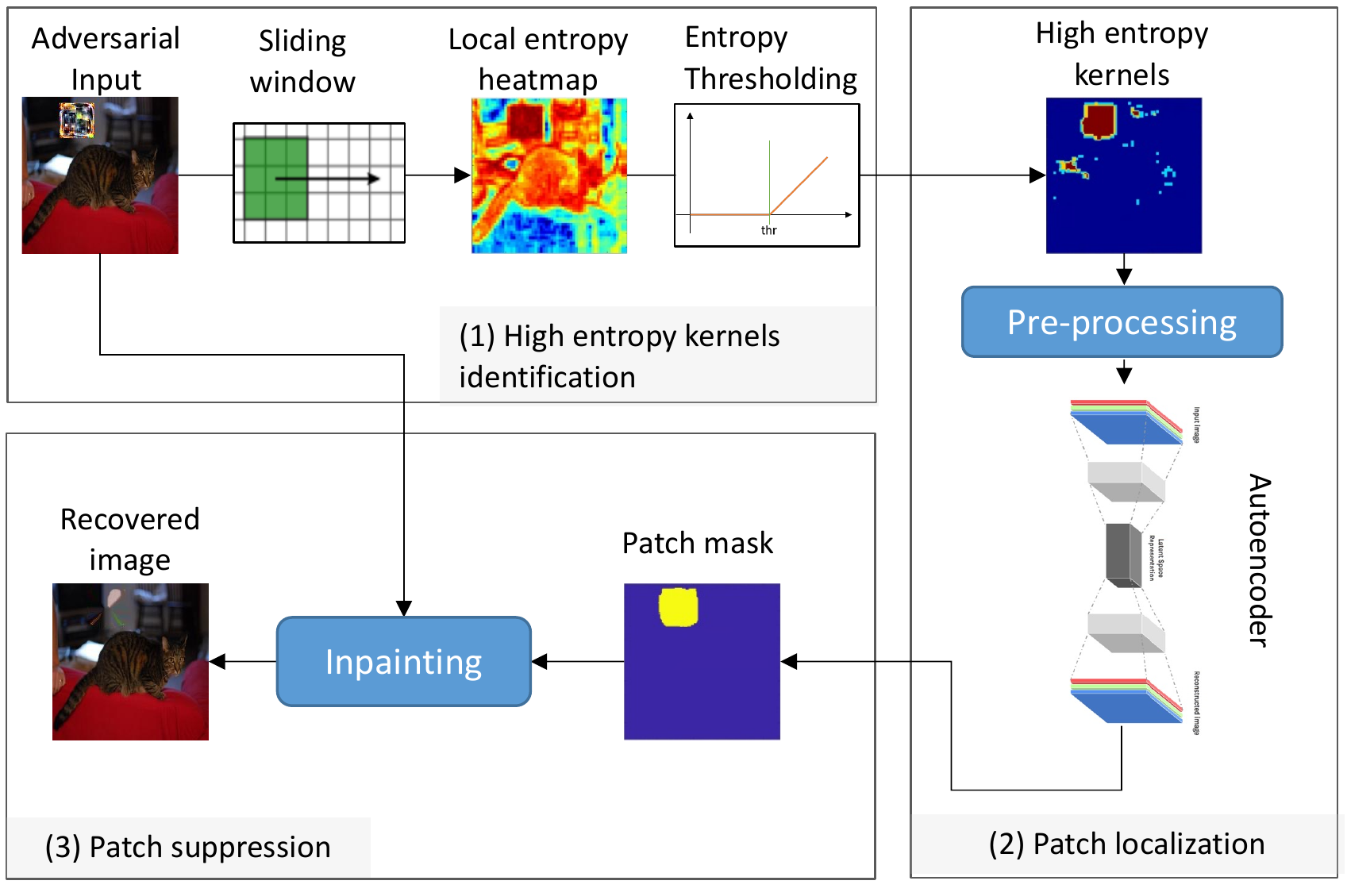}
  \caption{Detailed diagram of our Jedi adversarial patch defense}
  \label{fig_diag}
\end{figure}
Therefore, we propose to set a dynamic entropy threshold defined by Equation \ref{eq1}:
\begin{equation}
    thr = \mu_{clean} + (w_{tolerance} + w_{image}) \sigma_{clean}
    \label{eq1}
\end{equation}

Where: $\mu_{clean}$ and $\sigma_{clean}$ are, respectively, the mean and standard deviation of the clean samples entropy, $w_{tolerance}$ is an empirical weight to adjust the threshold according to the risk strategy, $w_{image}$ is the weight used to adjust the threshold according to the entropy distribution of the current image.
Finding the threshold requires a knowledge of the local entropy distributions: An entropy distribution of known clean images obtained from the considered dataset or application, and the entropy distribution of the current input image, easily extracted from the heat map. 

\noindent\textbf{Parameter exploration}: The  \emph{local entropy} exploration is based on the neighborhood definition, which depends on empirical spatial parameters such as the window size and the padding. Therefore, we proceed to a space exploration to select these parameters driven by maximizing the performance, which can be found in supplementary material. 

As shown in Figure \ref{fig_diag}, the outcome of this step is a heat map that contains high entropy "kernels" where potential patch candidates are detected. However, this kernel detection is not sufficient for the final mask localization as it may be incomplete and/or contain non-patch locations. Further processing is required to refine this map into the final mask.

\subsection{Patch shape reconstruction using Auto-Encoders}

The next step of building the mask to surgically localize adversarial patches is to expand the high entropy kernels to the full accurate shape of the patch using autoencoders (AEs) as these architectures are adapted to data reconstruction from incomplete inputs. 
The high entropy kernels obtained in the previous step may be incomplete or contain false positives; there may be some overlap between adversarial patch entropy values and some parts of the image with naturally high entropy. However, natural high entropy areas mostly correspond to edges, which results in scattered high entropy kernels. Therefore, to improve the quality of the final patch mask, we perform a pre-processing filtering where we remove all scattered clusters. 
We explored several AEs and chose the most effective architecture with the lowest latency. We use a sparse Autoencoder (SAE), which uses regularization to enforce sparsity. The proposed AE’s hyperparameters are as follows: (i) one input layer which takes the entropy heat map with the image size, (ii) one hidden layer of 100 neurons, and (iii) an output layer corresponding to the mask that identifies the patch location within the image. We use a sparse AE with a sparsity proportion of 0.15 and sparsity regularization of 4. To train the AE, we simulate an attack using a subset of images where we place a patch on the image, and  extract a mask representing the patch’s coverage; neither the images nor the patch were used other than in training the AE. Specifically, we collect these masks and use them as the training set of patch shapes for the AE to learn how to refine patch localization and provide a mask. 
We generated the AE training data by creating masks of the same size as the expected input images containing the most likely positions of adversarial patches and their potential sizes and shapes. The trained AE uses the high entropy kernel map as an input and outputs a mask with the reconstructed patch location. 
\subsection{Inpainting}
The last step of Jedi aims at recovering the prediction (detection/classification) lost due to the adversarial patch. Therefore, Jedi overwrites the localized patch with data sampled from the immediate neighborhood distribution using inpainting. 
Inpainting has been used by prior work \cite{jujutsu,dw,lance} to mitigate adversarial patch attacks given an accurate mask. 
Since the goal of this step is to defuse the patch, not to restore the exact original input, it is not required to have a pixel-perfect recovery of the original image. In fact, a sampling from the same distribution is sufficient for the current state-of-the-art DNN models to achieve correct output. 
We use the coherence transport based inpainting method \cite{inpaint_coherent}; Using the input image and the mask obtained from the previous step, we substitute mask pixels (starting from the boundaries and moving inwards) with a weighted sum of the values of external pixels within a certain radius. The weights are determined using the values of the coherence vectors in the pixel neighborhood.
 
\section{Experimental Methodology}
\label{sec:exp}

\subsection{Evaluation metrics}

We use the classical computer vision metrics (Accuracy for classification, Average Precision for detection) to evaluate the effectiveness of the attacks and our proposed defense's success in removing the patches. However, while these metrics can show a global view of the results of the patch attacks and defenses, they are not specific enough to evaluate and debug the defense mechanisms. For example, since \emph{patch detection} is fundamental for an accurate recovery, this needs to be evaluated separately. Moreover, the robust accuracy alone gives a global overview but does not allow a close examination of the defense behavior. In fact, if a given defense results in partially degrading the baseline accuracy and overall recovering, it is important to evaluate more precisely the impact. 
Therefore, we propose to additionally  consider the following set of metrics for a detailed analysis of the defense:

\noindent \textbf{i)} \textbf{ Patch Detection Rate:} It represents the percentage of applied patches that have been detected by the defense with an Intersection-over-union value that exceeds $0.5$. This metric enables a precise evaluation of the localization performance of a given defense. 

\noindent \textbf{ii)} \textbf{Recovery Rate:} The recovery rate represents the percentage of outputs restored by the defense, \emph{relative to the number of successful attacks.} It models precisely the intrinsic positive impact of the defense. 
    
   
\noindent \textbf{iii)}\textbf{ Lost Prediction Rate:} This metric models the performance degradation caused by the defense. It represents the percentage of negatively affected \textit{correct predictions} normalized to the set where the patch initially failed. 




\subsection{Experimental setup}



\noindent\textbf{Benchmarks:} We evaluate Jedi in a variety of environments commonly used in computer vision: \\
\noindent \textit{\underline{- Classification tasks:}} We use 
ImageNet \cite{imagenet} for the large amount and variety of classes it contains. We also use Pascal Dataset \cite{pascalvoc07} for the variety of entropy conditions: from very low background entropy (example: clear sky) to very high (urban environments and forests). 

\noindent \textit{\underline{- Detection tasks}:} for the two detection tasks, we use INRIA dataset \cite{inria} to test attacks in an outdoor high entropy uncontrolled environment, and CASIA datasets \cite{casia} to test detection in an indoor controlled environment.
 Other benchmarks could be found in the supplementary material.
 
\noindent\textbf{Adversarial Patches:} In this evaluation, we use four state-of-the-art adversarial patches: Adversarial Patch \cite{advpatch} and LAVAN  \cite{lavan} are used against the classification tasks. And the YOLO adversarial patch  \cite{yolopatch} and the Naturalistic Patch proposed \cite{naturalistic} are chosen to attack the detection tasks.


\noindent\textbf{Experiments:} For a comprehensive evaluation,  we chose combinations (model/dataset) \emph{where the generated patches were damaging}, which corresponds to a stronger attacker. This corresponds to ImageNet with \cite{advpatch}, using ResNet50, Pascal VOC 07 with \cite{advpatch}, using Resnet50, CASIA with \cite{yolopatch}, using YOLO and INRIA with \cite{yolopatch}. 
 Other combinations are available in the supplementary material.

\noindent\textbf{Comparison to the state-of-the-art:} We compare Jedi against LGS \cite{lgs} and Jujutsu \cite{jujutsu} in all of the experiments we describe earlier. We also for the sake of illustration compare with 3 certified defenses, namely, Derandomized Smoothing \cite{derandomized}, Smoothed ViT \cite{smoothed_vit} and Patchguard \cite{patchguard}.

\section{Results}
\label{sec:res}
\begin{figure}[tp]
    \centering
    \includegraphics[scale = 0.55]{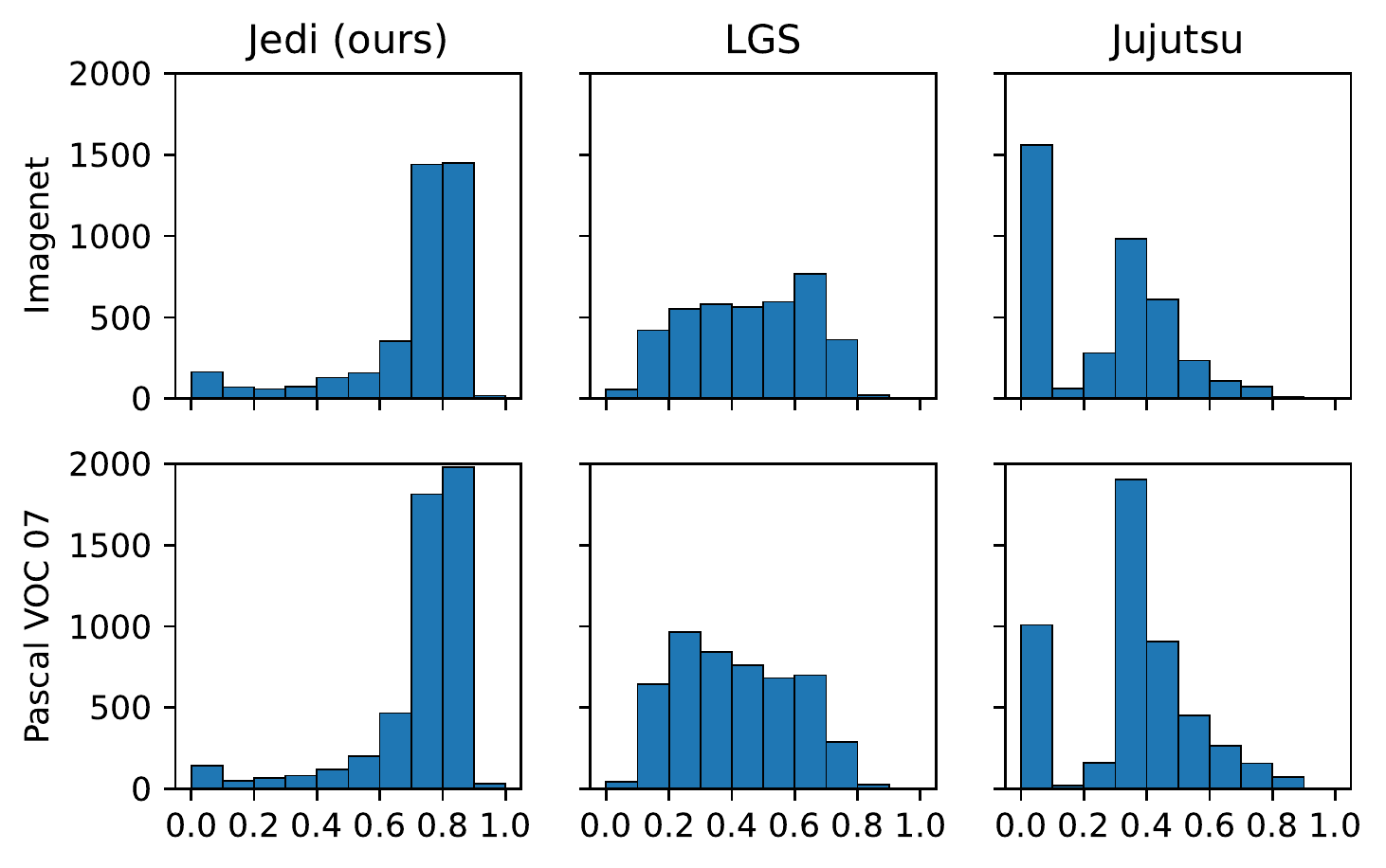}
    \caption{Comparison of detailed histograms of patch localization performance}
    \label{fig:loc_comp}
\end{figure}

\noindent\textbf{Patch localization performance -- }First, we focus on the patch localization performance; we measure the Intersection-over-Union (IoU) of the mask produced by Jedi, LGS and Jujutsu compared to the ground truth patch location and report the results in Table \ref{tab:tab_loc}. For a comprehensive qualitative study of the patch localization, we provide the comparative distribution of the IoU in Figure \ref{fig:loc_comp}, which depicts the occurrence of IoU scores in bins of $0.1$ width. The two rows of the figure show the two sample experiments considered for this qualitative study , on Imagenet and Pascal VOC 07, while the columns compare three patch mitigation methods: Jedi (ours, left), LGS (\cite{lgs},center) and Jujutsu (\cite{jujutsu}, right). 

\begin{table}
\centering
\caption{Patch localization performance }
\label{tab:tab_loc}
\begin{tabular}{@{}l|lll@{}}
\toprule
Dataset                         & \textbf{Jedi} (ours) & LGS     & Jujutsu \\ \midrule
Imagenet + \cite{lavan}         & 87.20\%    & 44.50\% & 10.85\% \\
Pascal VOC 07 + \cite{advpatch} & 90.71\%    & 34.30\% & 19.22\% \\
CASIA  + \cite{yolopatch}       & 93.49\%    & 57.25\% & N/A     \\
INRIA  + \cite{naturalistic}    & 38.80\%    & 73,11\% & N/A     \\ \bottomrule
\end{tabular}
\end{table}

An efficient patch localization process is characterized by a maximum coverage of the patch pixels and minimum false positive pixels, leading to high IoU values. Therefore, the IoU distribution of an efficient patch localization method skew towards high IoU. As shown in Figure \ref{fig:loc_comp}, Jedi has the highest shift to the right in the IoU distribution with more than $80\%$ of IoUs are higher than $0.7$, while less than $10\%$ of the two related defenses have IoUs in this range. 

\begin{figure}[tp]
    \centering
    \includegraphics[scale = 0.55]{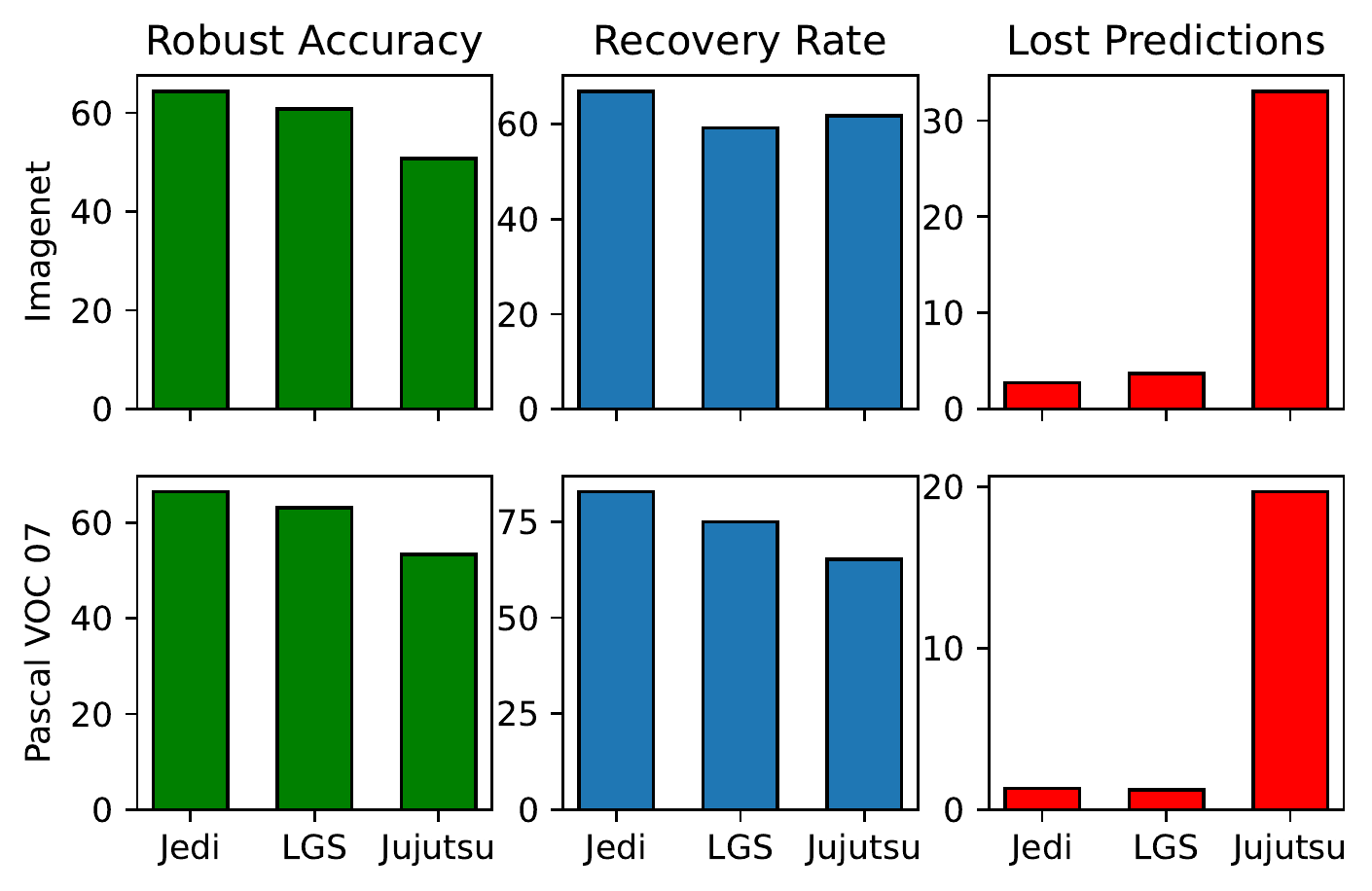}
    \caption{Comparison of Jedi's performance compared to other state of the art methods for classification tasks}
    \label{fig:comp_results_cls}
\end{figure}

\begin{figure}[htp]
    \centering
    \includegraphics[scale = 0.55]{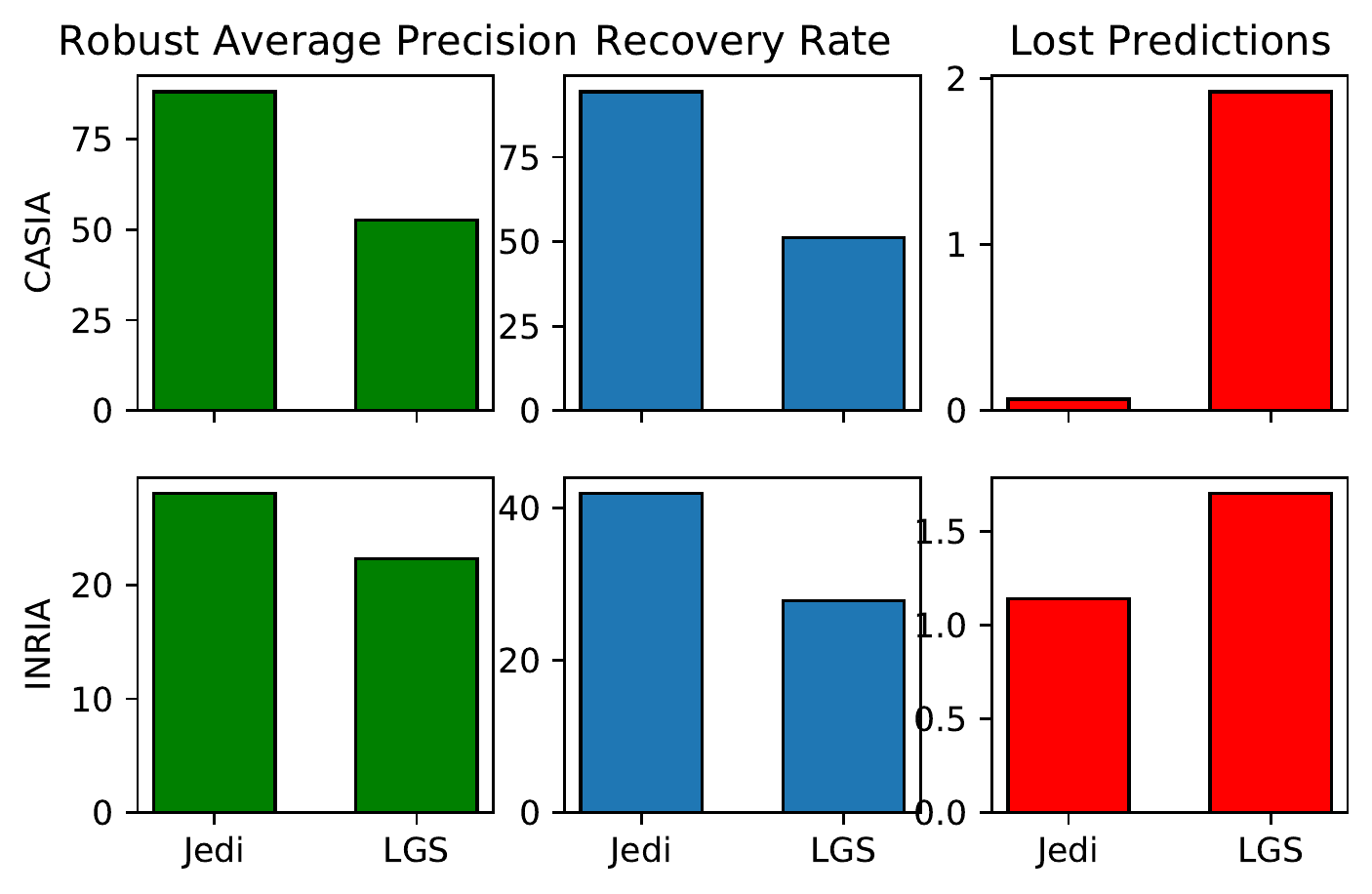}
    \caption{Comparison of Jedi's performance compared to other state of the art methods for detection tasks}
    \label{fig:comp_results_det}
\end{figure}

\noindent\textbf{End to end results -- }
In Table \ref{tab:tab_ap_cls}, we report the \emph{Clean Accuracy} of the model prior to adding the adversarial attack, \emph{Adversarial Accuracy} of the undefended model against the adversarial patch, and \emph{Robust Accuracy} of the model after applying Jedi in the classification tasks. Similarly, we report Clean Average Precision, Adversarial Average Precision and Robust Average Precision for the detection tasks in Table \ref{tab:tab_ap_det}. In Figure \ref{fig:comp_results_cls}, we report the Recovery Rate, Lost Predictions and Accuracy for Jedi for each of the adversarial datasets for the classification tasks, and compare the results to LGS and Jujutsu, while Figure \ref{fig:comp_results_det} depicts the comparative results in the detection tasks. Results show that Jedi outperforms related defenses for all metrics: Jedi restores most incorrect detections caused by adversarial patches while causing the lowest lost predictions. Further evaluation using different patch sizes is available in Section 6 of the supplementary material.

\begin{table}[tp]
\centering
\caption{Clean, Adversarial and Robust accuracy of Jedi in classification benchmarks}
\label{tab:tab_ap_cls}
\begin{tabular}{@{}l|lll@{}}
\toprule
Dataset       & \begin{tabular}[c]{@{}l@{}}Clean \\ Accuracy\end{tabular} 
              & \begin{tabular}[c]{@{}l@{}}Adversarial \\ Accuracy\end{tabular}
              & \begin{tabular}[c]{@{}l@{}}Robust \\ Accuracy\end{tabular} \\ \midrule
Imagenet      & 74.10\%        & 39.26\%              & 64.34\%         \\
Pascal VOC 07 & 72.17\%        & 26.94\%              & 66.40\%         \\ \bottomrule
\end{tabular}
\end{table}

\begin{table}[tp]
\centering
\caption{Clean, Adversarial and Robust accuracy of Jedi in detection benchmarks}
\label{tab:tab_ap_det}
\begin{tabular}{@{}l|lll@{}}
\toprule
Dataset &
  \begin{tabular}[c]{@{}l@{}}Clean Avg \\ Precision\end{tabular} &
  \begin{tabular}[c]{@{}l@{}}Adversarial Avg\\ Precision\end{tabular} &
  \begin{tabular}[c]{@{}l@{}}Robust Avg \\ Precision\end{tabular} \\ \midrule
CASIA &
  91.47\% &
  39.60\% &
  88.21\% \\
INRIA &
  53.22\% &
  12.17\% &
  28.03\% \\ \bottomrule
\end{tabular}
\end{table}

 We also evaluate certified defenses on Imagenet dataset, using \cite{advpatch} and Resnet-50, the baseline results using an unprotected model are Clean Accuracy of $74.10\%$, the Patch Success rate is $49.02\%$ and an Adversarial Accuracy of $39.26\%$. While a direct comparison is unfair, we wish to quantify the cost of ensuring provable robust defense on overall performance. As expected, their performance was lower than the empirical defenses, as shown in Table \ref{tab:res_cert}. These results show that while certified adversarial patch defenses offer provable robustness, their empirical utility is limited. These issues are more evident when the input images are larger or contain more aggressive attacks (such as bigger patches). 
 

\begin{table}[]
\centering
\caption{Performance of certified defenses on the Imagenet dataset}
\label{tab:res_cert}
\begin{tabular}{@{}l|llll@{}}
\toprule
 Defense & Jedi & \cite{derandomized} & \cite{patchguard} & \cite{smoothed_vit}\\ \midrule
Robust Accuracy      & \textbf{64.34}\% & 35.02\% & 30.96\% & 40.38\% \\
Recovery Rate        & \textbf{66.95}\% & 27.65\% & 28.07\% & 36.34\% \\
Lost Predictions     & \textbf{2.77}\%  & 52.56\% & 51.55\% & 33.54\% \\ \bottomrule
\end{tabular}
\end{table}

\section{Adaptive Attacks}
\label{sec:adapt}
Empirical defenses against adversarial attacks have been shown vulnerable to adaptive attacks, where an attacker is aware of the defense and its parameters (i.e. a white box scenario). The adversary creates an adversarial patch that exploits specific weakness in the defense to bypass it. 
In this section, we investigate the robustness of Jedi to 2 adaptive attacks: (i) the first is a GAN-based naturalistic patch generation method, (ii) and the second is an \textbf{entropy-aware adaptive attack} that we propose.

\subsection{Naturalistic Patch}

\begin{figure}[h]
     \centering
     \begin{subfigure}[b]{0.15\textwidth}
         \centering
         \includegraphics[width=0.7\textwidth]{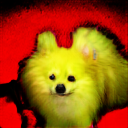}
         \caption{CASIA dataset}
         \label{fig:ihr}
     \end{subfigure}
     \begin{subfigure}[b]{0.15\textwidth}
         \centering
         \includegraphics[width=0.7\textwidth]{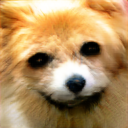}
         \caption{INRIA dataset}
         \label{fig:cv}
     \end{subfigure}
        \caption{Samples of the naturalistic patches}
        \label{fig_nat_patch}
\end{figure}

A recent adversarial patch generation method that might be adaptive to our defense is the Naturalistic Patch \cite{naturalistic}. 
We investigate the effectiveness of this attack since it generates stealthy adversarial patches that mimic real objects to avoid visual suspicion and therefore has the potential to evade detection. Samples of the naturalistic patch generated for our experiments are shown in Figure \ref{fig_nat_patch}.

\begin{figure}[tp]
    \centering
    \includegraphics[width=\columnwidth]{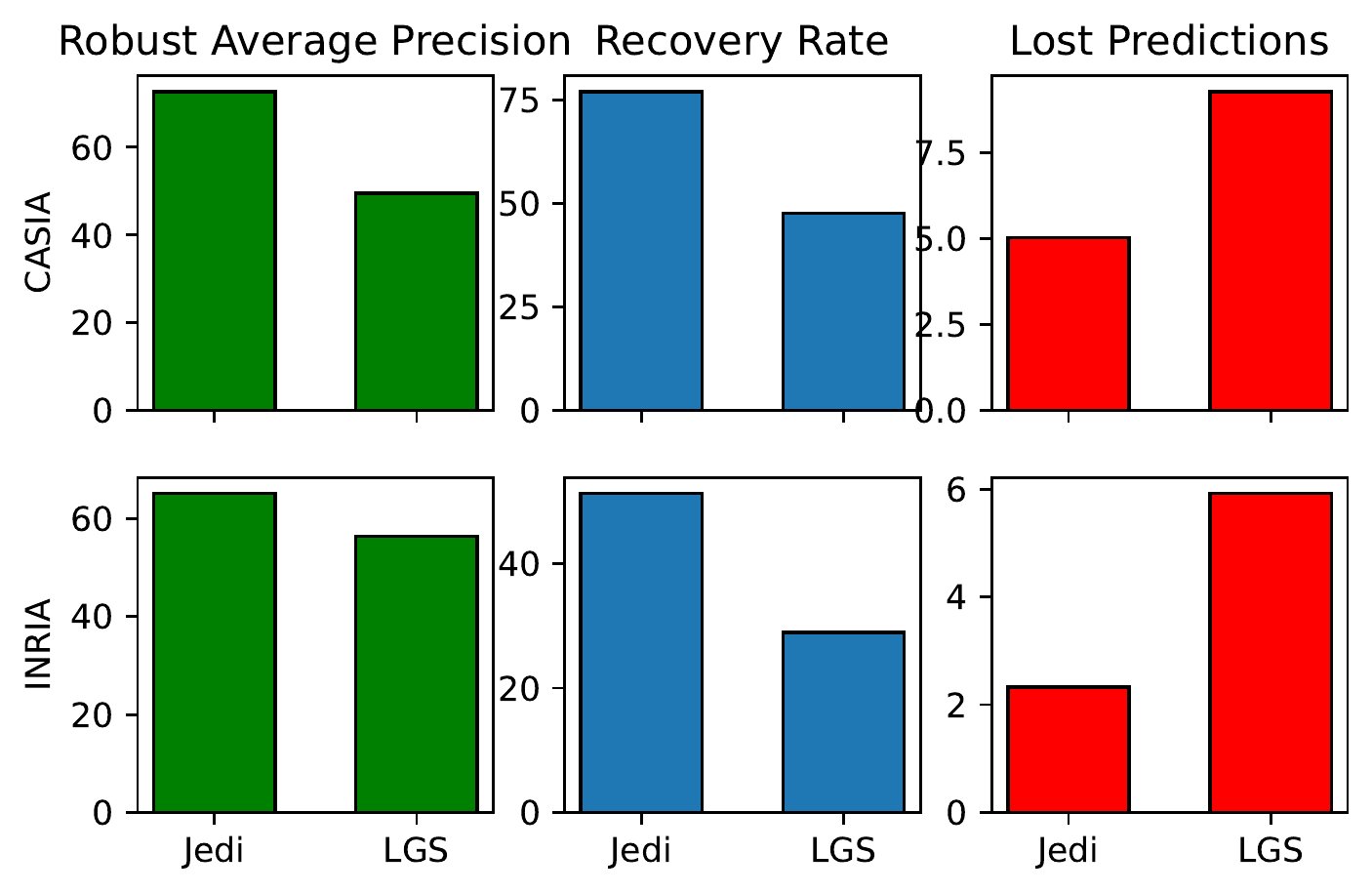}
    \caption{Comparison of Jedi's performance against the naturalistic patch}
    \label{fig:comp_results_nat}
\end{figure}

The assumption is that these naturalistic patches are outliers to our initial observation in terms of entropy distribution (Section 2). Therefore, one can expect these patches to evade our defense. We evaluate Jedi against the naturalistic patches and we summarize the results in Figure \ref{fig:comp_results_nat}. Please note that we found that the naturalistic patch wasn't efficient in classification tasks (patch success rate didn't exceed $8\%$). For this reason, we show only detection benchmarks results.   
Surprisingly, the results indicate that Jedi is able to detect and mitigate the naturalistic patches with similar performance to regular patches, while LGS efficiency is considerably impacted by the Naturalistic Patch. The takeaway from this experiment is that adversarial patches, perhaps inherently, require high entropy even if they look naturalistic.
In the following, we further investigate this hypothesis by generating an \emph{entropy-bounded adversarial patch}.

\subsection{Low-entropy adaptive patch}

To create a patch that bypasses our defense, it should contain a low entropy by-design. Therefore, we propose an adversarial patch generation method with an objective to limit the patch entropy. However, since Shannon's entropy is not derivable we define a constraint on entropy instead of integrating it in the loss function. 
Specifically, we formulate the problem as a constrained optimization; Our objective is to fool a victim model $C(.)$ on almost all the input samples from a distribution $\mu$ in $\mathbb{R}^d$ given an entropy budget $\varepsilon$. This problem can be expressed as finding $\delta$ such that:
\begin{equation}\label{eq:srp}
\begin{split}
        C(x + \mathcal{A}(\delta, k)) \neq C(x), ~ for ~ most ~ x \sim \mu \\ s.t. \mathcal{H}(\delta) \leq \varepsilon 
\end{split}
\end{equation}

Where $\delta$ is the adversarial patch, which entropy is controlled by $\varepsilon$, $\mathcal{A}(.)$ is a function that applies a patch $\delta$ in a given location $k$ within an input sample $x$.

\begin{figure}[tp]
    \centering
    \includegraphics[width=\columnwidth]{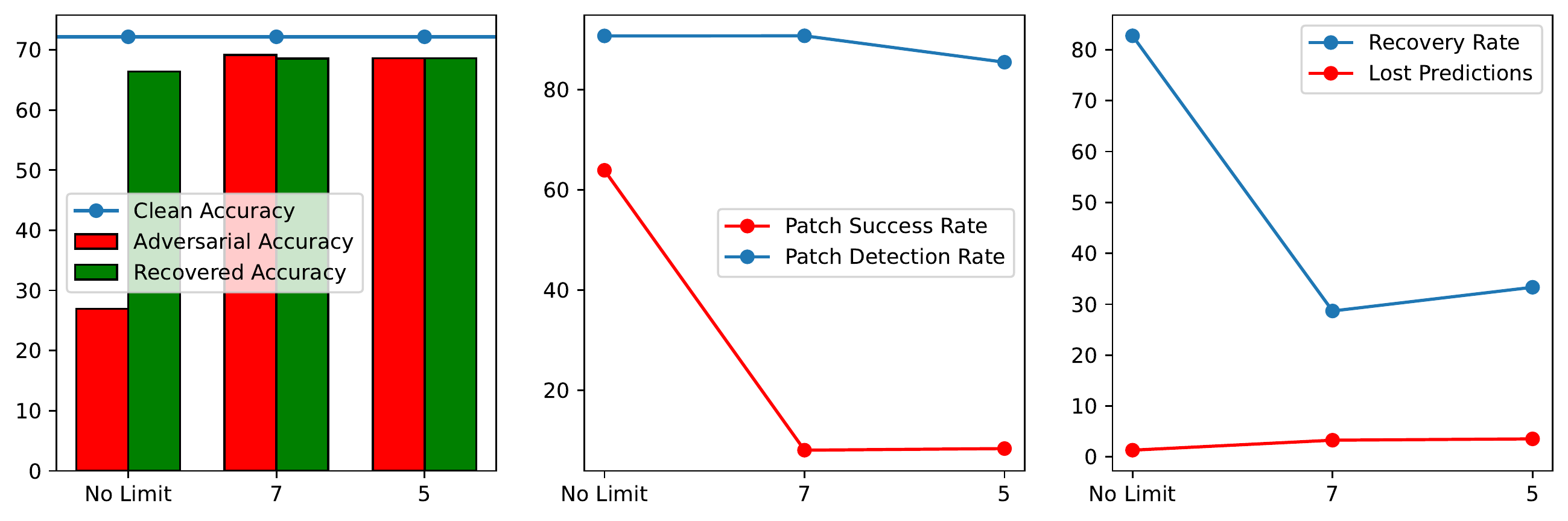}
    \caption{Performance of low entropy adaptive patches with comparison to a regular patch}
    \label{fig:adapt_perf}
\end{figure}

The patch generation approach is shown in Algorithm \ref{alg1}; To enforce the entropy constraint, we need to project the noise $\delta$ back to the entropy-limited space. Therefore, we add a parallel process that halts the patch generation every $n$ iterations to check whether the patch satisfies the entropy constraint. If not, a  \emph{local search for entropy reduction} is run to project the patch in the acceptable solutions space. This step is based on Variable Neighborhood Search, where for each iteration of entropy reduction, a search is initiated to find the best neighboring patch by replacing pixels of a certain color value by the closest (euclidean distance in the RGB space) color that already exists in the image. This step reduces the number of colors in the image, which in turn reduces the entropy. The selected neighbor is the patch that keeps the highest attack success rate. This process is repeated until patch entropy is below the budget, then the patch generation process resumes. Using this process, we attempt to create a patch that bypasses our defense on Pascal VOC 07 classification task. The results are shown in Figure \ref{fig:adapt_perf} for patches with different entropy budgets, as well as a regular patch for reference. 
Figure \ref{fig:adapt_perf} shows that the patches with a lower entropy limit can partially bypass Jedi, with lower patch detection rates and recovery rates. However, the low entropy patch has lost nearly all of its capabilities: The attack success rate drops from $64\%$ down to only $8\%$.
The source code as well as illustrations of the entropy limit's impact on the generated patches are available in the supplementary materials. 


\begin{algorithm}
    \caption{Low entropy patch generation algorithm}
    \label{alg1}
    \begin{algorithmic}[1]
        \State \textbf{Input:}{ $n_{epochs}$: number of training epochs ,$I_{train}$ Training image set,$params$: Patch generation parameters, $checkFreq$: Frequency of entropy checks , $\varepsilon$: entropy limit}
        \State \textbf{Output:}{ Low entropy adversarial patch}
        \For {$e \in [1,n_{epochs}]$}
            \For {$im \in I_{train}$}
                \State $\delta$ = \textit{PatchTraining}(params)
                \If{$it \% checkFreq == 0$ }
                \State \textit{/*Repeat each \textit{checkFreq} iterations*/}
                    \State patchEntropy $\xleftarrow{}$ \textit{Entropy}($\delta$)
                    \While{$patchEntropy > \varepsilon$}
                        \State \textit{/*Target n random colors in the image for entropy reduction*/}
                        \State colorList $\xleftarrow{}$ \textit{random}(0,255,n)
                        \State $\delta$ $\xleftarrow{}$ reduceEntropy($\delta$,colorList)
                    \EndWhile
                    \State $patchEntropy$ $\xleftarrow{}$ \textit{entropy}($\delta$)
                \EndIf
            \EndFor
        \EndFor
        \State
        \State \textbf{function} \textit{reduceEntropy}($\delta$, colorList)
        \State \textit{/*Find best patch among random color removal tries*/}
        \For{$c \in colorList$}
        \State \textit{ /*Remove the targeted color and replace it with the most similar color found in the image*/}
        \State nearestColor $\xleftarrow{}$ 
        \textit{findNearestColor}(c)
        \State newPatch[c] $\xleftarrow{}$ \textit{replaceColor}($\delta$, c, nearestColor)
        \State newPatchASR[c] $\xleftarrow{}$ \textit{testSuccessRate}(newPatch[c])
        \EndFor
        \State \textit{ /*Keep the best performing patch */}
        \State bestPatch $\xleftarrow{}$ \textit{argmax}(newPatchASR)
        \State \textbf{return} bestPatch
        \State \textbf{end function}
    \end{algorithmic}
\end{algorithm}
 
\begin{figure}[tp]
  \centering
  \includegraphics[width=\columnwidth]{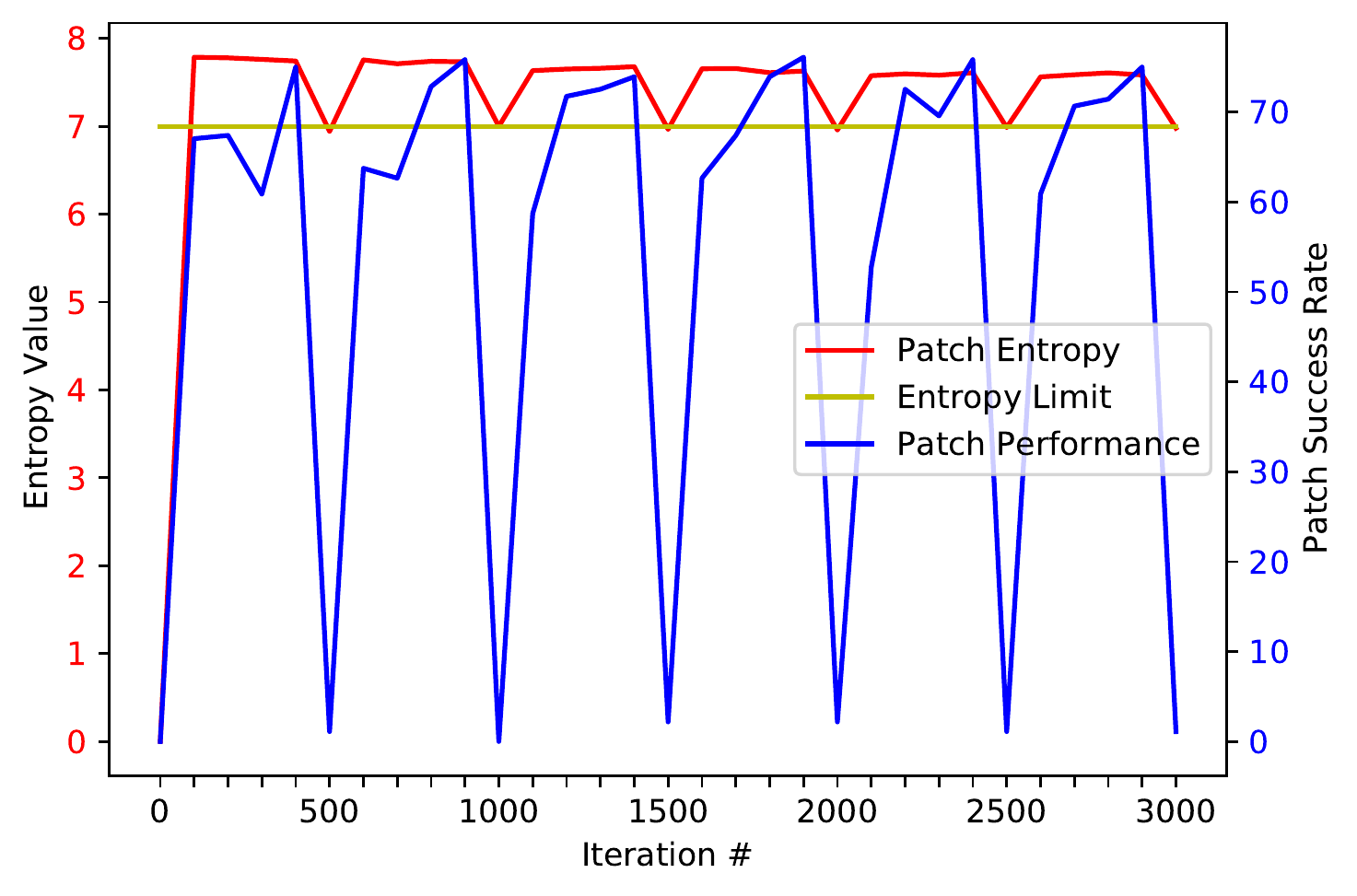}
  \caption{Entropy evolution during the adaptive patch generation}
  \label{fig_adapt_sample}
\end{figure}

\section{Discussion and Concluding Remarks}
\label{sec:disc}


In this paper, we present Jedi, a new defense against adversarial patches that is model agnostic, and robust against naturalistic patches. To our knowledge, this work is first to leverage a \textbf{differential entropy analysis} to detect adversarial patches. For a surgical localization, the entropy exploration outcome is fed to an AE which generates patches that are then overwritten by surrounding distribution through an inpainting.  
Jedi accurately localizes $85.3\%$-$93.5\%$ of adversarial patches across different datasets/attacks. Our qualitative analysis shows that the majority of detected patches have high IoUs. This confirms that \emph{entropy distribution is a reliable metric to separate adversarial patches from benign images}. Given the efficient patch localization, the end-to-end post-inpainting results show accuracy/average precision restored up to their clean level. Across several benchmarks, Jedi recovers $67.0\%$-$94.4\%$ of the originally correct detections, with a low lost predictions compared to related work.

\noindent\textbf{Comparison with certified defenses.} To compare with certified defenses, we tested Patch Cleanser \cite{patchcleanser} comparatively to Jedi. For Imagenet benchmark with Resnet50, Patch Cleanser with 2x2 mask grid achieved $56.97\%$ non certified robust accuracy, compared to $64.36\%$ for Jedi. The certified accuracy for this benchmark is $4.87\%$. For this setting Jedi offers higher robustness under comparable time per frame. For a 6x6 mask grid, Patch Cleanser achieves $64.00\%$ robust accuracy, but consumes $\sim 10 \times$ more time than Jedi.

\noindent\textbf{Adaptive attacks.} We tested Jedi against an attack that uses GANs to generate naturalistic patches \cite{naturalistic}, which a priori seemed adaptive to our defense since it is based on distinguishing  distribution of natural images from adversarial noise distribution. Interestingly, Jedi shows high robustness even against this attack. To further evaluate the vulnerability against potential adaptive attacks, we attempted to exploit what seems to be a weak spot for an adversary; A low entropy patch might go undetected by our defense. 
Our experiments show that to generate a Jedi-undetectable patch, the adversary has to sacrifice the attack efficiency, rendering the adaptive patch useless from an attacker perspective. On the other hand, rising the entropy budget increases the patch efficiency but accordingly makes it within Jedi detection capacity. 

\noindent\textbf{Limits. }We believe that these adaptive attacks corroborate our initial intuition that adversarial patches might \textbf{inherently} require high entropy due to the information theoretical necessity of embedding too much information in a limited "channel". While this is based on empirical analysis, we believe that further analysis from information theoretical perspective is required to prove these findings.



\section*{Acknowledgment}
This work has been supported in part by RESIST project funded by Région Hauts-de-France through STIMULE scheme (AR 21006614), and EdgeAI KDT-JU European project (101097300).

{\small
\bibliographystyle{ieee_fullname}
\bibliography{bib}
}


\end{document}